\documentclass[conference]{IEEEtran}
\IEEEoverridecommandlockouts
\usepackage{cite}
\usepackage[numbers]{natbib}
\usepackage{amsmath,graphicx,booktabs,arydshln}
\usepackage{amsmath,amssymb,amsfonts}
\usepackage{algorithmic}
\usepackage{graphicx}
\usepackage{textcomp}
\usepackage{url}
\usepackage{xcolor}
\def\BibTeX{{\rm B\kern-.05em{\sc i\kern-.025em b}\kern-.08em
    T\kern-.1667em\lower.7ex\hbox{E}\kern-.125emX}}
\begin{document}

\title{The Impact of Audio Input Representations on Neural Network based Music Transcription\\
}

\author{\IEEEauthorblockN{Kin Wai Cheuk}
\IEEEauthorblockA{\textit{Information Systems,}\\
	\textit{Technology, and Design}\\
	\textit{Singapore University}\\
	\textit{of Technology and Design}\\\\
	\textit{Institute of}\\
	\textit{High Performance Computing}\\
	\textit{Agency for Science,}\\
	\textit{Technology and Research} \\
Singapore\\
kinwai\_cheuk@mymail.sutd.edu.sg}
\and
\IEEEauthorblockN{Kat Agres}
\IEEEauthorblockA{\textit{Yong Siew Toh Conservatory of Music} \\
	\textit{National University of Singapore}\\
	\textit{Institute of}\\
	\textit{High Performance Computing}\\
	\textit{Agency for Science,}\\
	\textit{Technology and Research} \\
Singapore \\
muskra@nus.edu.sg}
\and
\IEEEauthorblockN{Dorien Herremans}
\IEEEauthorblockA{\textit{Information Systems,}\\
	\textit{Technology, and Design}\\
	\textit{Singapore University}\\
	\textit{of Technology and Design}\\\\
	\textit{Institute of}\\
	\textit{High Performance Computing}\\
	\textit{Agency for Science,}\\
	\textit{Technology and Research} \\
	Singapore\\
dorien\_herremans@sutd.edu.sg}
}

\maketitle

\begin{abstract}
This paper thoroughly analyses the effect of different input representations on polyphonic multi-instrument music transcription. We use our own GPU based spectrogram extraction tool, nnAudio, to investigate the influence of using a linear-frequency spectrogram, log-frequency spectrogram, Mel spectrogram, and constant-Q transform (CQT). Our results show that a $8.33$\% increase in transcription accuracy and a $9.39$\% reduction in error can be obtained by choosing the appropriate input representation (log-frequency spectrogram with STFT window length 4,096 and 2,048 frequency bins in the spectrogram) without changing the neural network design (single layer fully connected). Our experiments also show that Mel spectrogram is a compact representation for which we can reduce the number of frequency bins to only 512 while still keeping a relatively high music transcription accuracy.
\end{abstract}

\begin{IEEEkeywords}
Automatic Music Transcription, Spectrogram, Neural Network, Audio input representation
\end{IEEEkeywords}

\section{Introduction}
\label{sec:intro}
Polyphonic music transcription is extremely difficult, yet it is a fundamental step to other music information retrieval tasks~\cite{Benetos2019AutomaticMT, Klapuri2006,Benetos2013AutomaticMT, BoulangerLewandowski2012ModelingTD}. Most automatic music transcription (AMT) research focuses on developing sophisticated models for the transcription problem. Various models such as support vector machines (SVM)~\cite{Poliner2006ADM,Poliner2007IMPROVINGGF}, restricted Boltzmann machines (RBM)~\cite{BoulangerLewandowski2012ModelingTD}, long-short term memory neural networks~\cite{Bck2012PolyphonicPN}, and convolutional neural networks (CNN)~\cite{Thickstun2017InvariancesAD, Trabelsi2017DeepCN} have been developed to tackle this task. For example, \citet{Wang2017ATA} integrate non-negative matrix factorization (NMF) with a CNN in order to improve transcription accuracy. \citet{Hawthorne2017OnsetsAF} split the AMT into three sub-tasks: onset detection, frame activation, and velocity estimation, which allows them to achieve state-of-the art transcription accuracy on piano music. Most literature focuses mainly on polyphonic piano transcription and their models are being trained on piano datasets such as MAPS~\cite{emiya2010maps}. The existing studies on music transcription often use different input representations such as log magnitude spectrogram~\cite{Poliner2007IMPROVINGGF, Poliner2006ADM}, constant-q transform (CQT)~\cite{Wang2017ATA, Bck2012PolyphonicPN}, and Mel spectrogram~\cite{Hawthorne2017OnsetsAF}. Only few studies, however, offer a comparison of the effect of different input representations~\cite{Thickstun2017InvariancesAD,Logan2000MelFC, Kelz2016OnTP, balamurali2019toward}.
One of the most comprehensive studies~\cite{Kelz2016OnTP} compares the effect of using spectrograms with linearly and logarithmically spaced frequency bins versus CQT. Their results show that a single layer, fully connected network that uses logarithmic frequency bins with a logarithmic magnitude spectrogram performs best on polyphonic piano transcription with the MAPS dataset. There are, however, several missing aspects in their study. First, a comparison with one of the popular representations, Mel spectrogram, is missing. Second, the effect of the number of bins for different representations was not examined. Finally, they did not test whether the same approach may be applied for polyphonic multi-instrument transcription.

Our study provides a more comprehensive analysis of the effect that input representation has on polyphonic audio transcription accuracy for multiple instruments when using a single layer fully connected network. We make use of our own GPU based spectrogram toolkit in Python (nnAudio)~\cite{cheuk2019nnaudio, nnAudio} to allow for on-the-fly spectrogram calculations. The main contributions of this research are to: 1) provide a comprehensive analysis of the effect of four popular frequency domain input representations on polyphonic music transcription; 2) provide an analysis of model performance when varying the number of bins for each of the input representations; and 3) study the model's performance when varying the input frame length. We first discuss our base model for transcription and input representations before moving on to the experimental setup and results. Finally, a conclusion wraps up the paper together with suggestions for future work. The source code is available at our github page\footnote{\url{https://github.com/KinWaiCheuk/IJCNN2020_music_transcription}.}. 

\section{System description}
\begin{figure*}[h!]
	\centering
	\includegraphics[width=\linewidth]{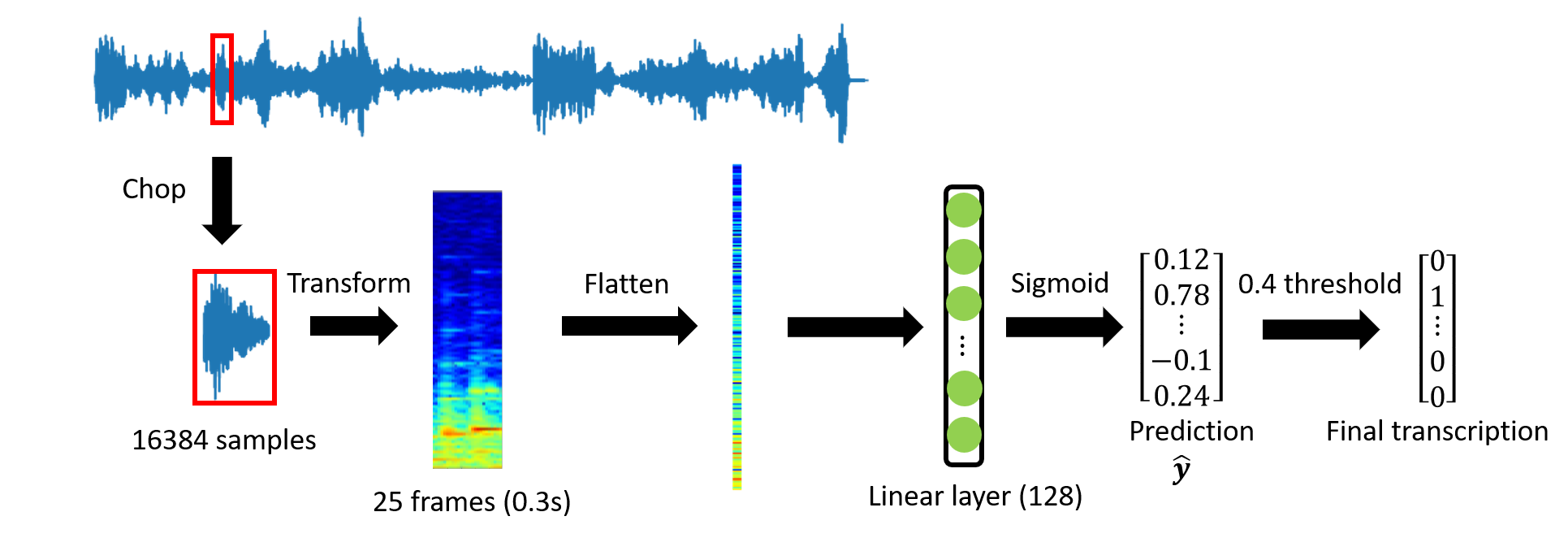}
	\caption{The MusicNet model from~\citet{Thickstun2017InvariancesAD} is used as our base architecture.}
	\label{fig:Linear}
\end{figure*}
\subsection{Input Representations}
\begin{figure}[htb]
	\centering
	\includegraphics[width=9.0cm]{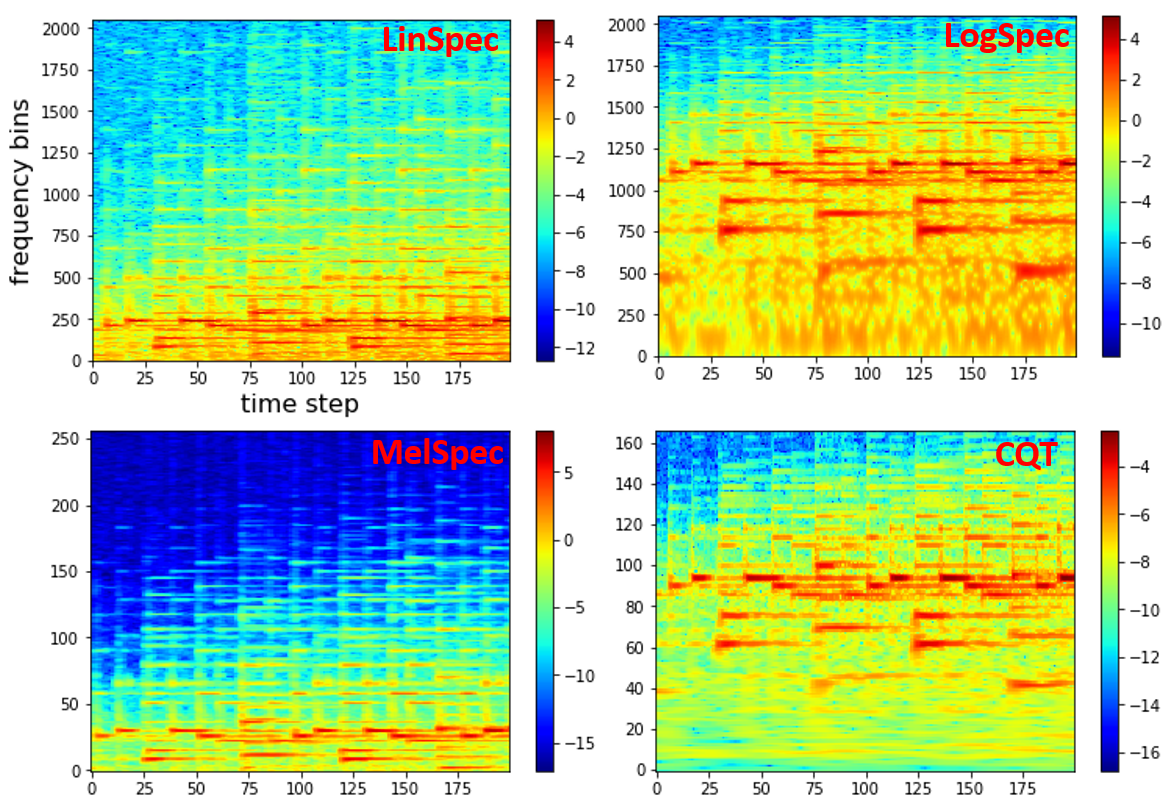}
	\caption{Excerpt from Bach's WTK I, No. 5, Prelude (ID:2303) visualized by four input representations: linear-frequency spectrogram, log-frequency spectrogram, Mel spectrogram, and CQT (clockwise from top left).}
	\label{fig:input_respresentation}
\end{figure}

Our experiments compare the influence of four different input representations: linear-frequency spectrogram (LinSpec), log-frequency spectrogram (LogSpec), CQT, and Mel spectrogram (MelSpec). In order to study the effect of different spectrogram parameters, we developed a GPU based spectrogram conversion toolkit using Pytorch, called nnAudio\footnote{\url{https://github.com/KinWaiCheuk/nnAudio}}. This allows us to generate different types of spectrograms on-the-fly while training the network. Fig.~\ref{fig:input_respresentation} shows the four input representations studied in this paper. The sampling rate of the audio clips follows the work of~\citet{Thickstun2016LearningFO, Thickstun2017InvariancesAD}, which are kept at $44,100$Hz, and the hop size for waveforms to spectrograms transformation is kept constant at $512$ throughout our experiments.

For the sake of convenience, we will briefly mention the mathematics behind each spectrogram representation and the differences among them to save readers troubles from looking them up individually. 

\subsubsection{LinSpec}
This input type is the typical spectrogram obtained with short-time Fourier transform (STFT) as shown in equation~\ref{STFT}, where the $k$ inside $X[k,\tau]$ is the frequency bins of the spectrogram and the $\tau$ inside  $X[k,\tau]$ is the time-step of the spectrogram. $N$ is the window size, $n$ is the sample index for the input waveform, and $h$ is the hop size. Each frequency bin $k$ corresponds to a physical frequency $f$ in the unit of Hertz through the relationship of $ f= k\frac{s}{N} \label{f-k-conversion}$, where $s$ is the sampling rate. In this case, the spacing between each $k$ is linear, and hence this spectrogram has a linear frequency along the y-axis.

\begin{equation}
	X[k,\tau] = \sum_{n=0}^{N-1}x[n+h\tau]\cdot e^{-2\pi i k \frac{n}{N}}.\label{STFT}
\end{equation}

\subsubsection{LogSpec}
This spectrogram shares the same equation~\ref{STFT} as LinSpec, except the spacing between frequency bin $k$ is logarithmic instead of linear. Since the musical notes have a logarithmic relationship with each other, one would expect that LogSpec works well for music.

\subsubsection{MelSpec}
\begin{figure}[htb]
	\centering
	\includegraphics[width=9.0cm]{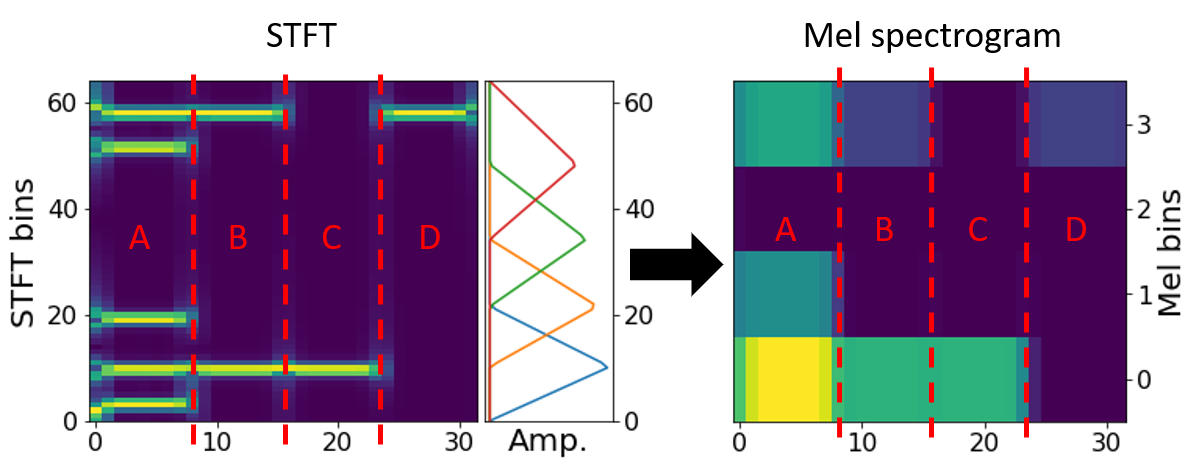}
	\caption{Mel filter banks can be used to convert a STFT output to Mel spectrogram. Each Mel filter bank covers multiple STFT bins $k$ and reduces them into one Mel bin.}
	\label{fig:mel_filter_banks}
\end{figure}

Mel spectrograms use a Mel scale which is an attempt to imitate humans' perceptual hearing on musical pitches~\cite{Stevens1937ASF, umesh1999fitting, o1987speech}. There are various versions of Mel scale, in this paper, we use the version implemented in the HTK Speech Recognition toolkit~\cite{young2002htk}. Then, we can use Mel filter banks to map frequency bins $k$ from STFT to Mel bins in Mel spectrograms as shown in Fig.~\ref{fig:mel_filter_banks}.

\subsubsection{CQT}
Constant-Q transformation (CQT)~\cite{youngberg1978constant, brown1991calculation, brown1992efficient} is a modification of the STFT with the formula as shown in equation~\ref{CQT}.

\begin{equation}
X^{cq}[k_{cq},\tau]= \sum_{n=0}^{N_{k_{cq}}-1}x[n+h\tau]\cdot e^{-2\pi i Q \frac{n}{N_{k_{cq}}}}
\label{CQT}
\end{equation}

Different from STFT, CQT has a constant factor $Q$ instead of the variable $k$ in $e^{-2\pi i Q \frac{n}{N_{k_{cq}}}}$. Also, the window length $N_{k_{cq}}$ for CQT is now a variable instead of a constant. The constant factor $Q$ is defined as $Q=(2^{\frac{1}{b}}-1)^{-1}$ and the variable window length is defined as $N_{k_{cq}}=   \text{ceil}{\left(\frac{s}{f_{k_{cq}}}\right)Q}$. Here $b$ is the number of bins per octave, which is a parameter for CQT.

\subsection{Model Architecture}

We use the architecture proposed by \citet{Thickstun2017InvariancesAD} as our base model (see Fig.~\ref{fig:Linear}), with which we test the influence of different spectrogram representations. This architecture consists of only one linearly fully connected layer with 128 neurons, whereby the spectrograms are flattened before feeding them as input to the model. The neural network architecture is kept minimal in order to focus the effect of input representations on music transcription accuracy. The network is trained for 35 epochs, in batches of size 1,000. Each batch contains 100 spectrograms sampled from the dataset. Adam optimizer with learning rate $10^{-6}$ is used due to a faster convergence. The output $\hat{y}$ of the neural network is used to calculate the binary cross entropy with the pitch label obtained at the time step that corresponds to the the center of the spectrogram (red dotted line in Fig.~\ref{fig:Linear}).

\section{Experimental Setup}

\subsection{Dataset and evaluation}
We perform a number of experiments using the MusicNet~\cite{Thickstun2016LearningFO} dataset. The dataset contains 330 classical music recordings with multi-pitch annotations aligned using dynamic time warping, and verified by trained musicians. Unlike previous datasets, which focus mostly on piano music~\cite{emiya2010maps}, MusicNet contains various musical instruments such as violin, clarinet, flute, and even harpsichord. This makes the transcription task much harder and realistic. The training set consists of 320 recordings, and the remaining ten recordings are used for evaluation. We use the same split for training and evaluation as \citet{Thickstun2017InvariancesAD}, and use precision, accuracy, and error as defined by \textbf{mir\_eval.multipitch.metrics}~\cite{Raffel2014MIR_EVALAT}, to ensure a fair comparison. 

\subsection{Experiment 1: Resolution of Spectrograms}\label{sec:Exp1}
In this experiment, we study the effect of \emph{varying the number of spectrogram bins} from roughly 256 to 2,048 for each of the input representations. The bin spacings have been carefully chosen such that the first bin starts at $50$Hz and the last bin ends at $6,000$Hz for all of the input representations. For LinSpec and LogSpec, the short-time Fourier transform (STFT) window size is fixed to 4,096, and the number of frequency bins is varied from 256 to 2,048. A smaller number of frequency bins means a larger frequency spacing between each bin. In other words, the spectrogram resolution is changed by the frequency bins. 
\begin{figure}[htb]
	\centering
	\includegraphics[width=8.5cm]{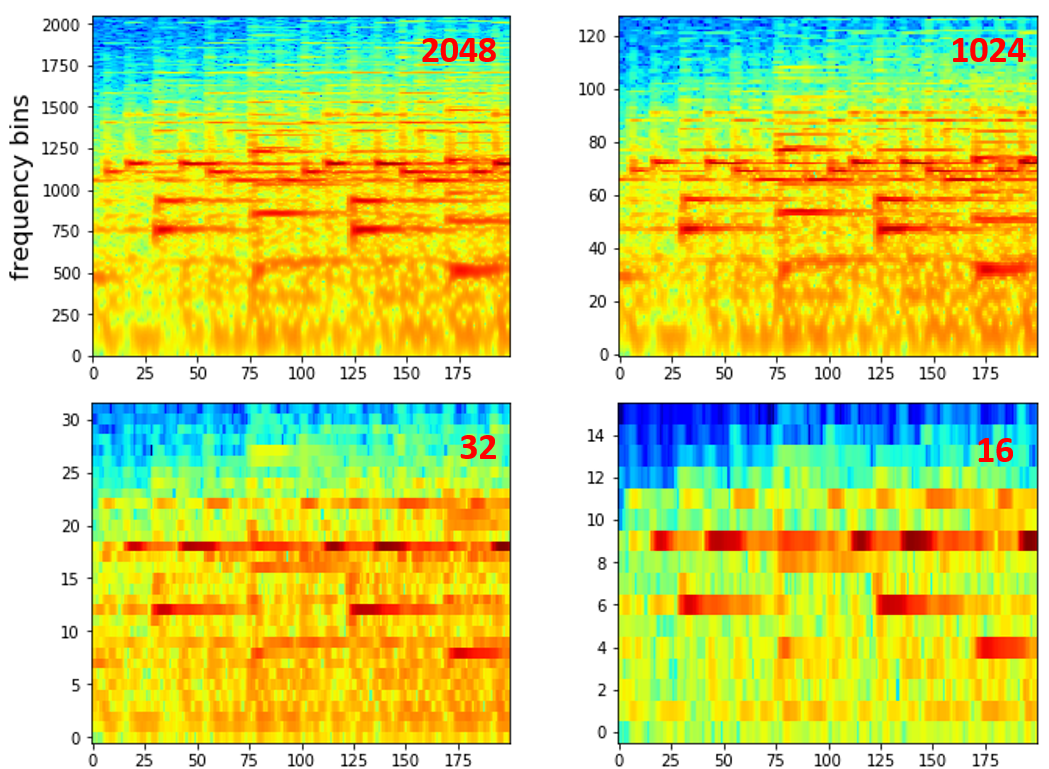}
	\caption{LogSpec of song ID:2303 with varying bin size of (2,048, 1,024, 32, and 16).}
	\label{fig:Exp1}
\end{figure}

For CQT, the first bin corresponds to the musical note A1 and the last bin corresponds to F$\sharp$8 (encompassing 84 notes from $55$Hz to $5919$Hz). During CQT calculation, the number of bins of the spectrogram can be tuned by adjusting the number of bins per octave. If the number of bins per octave is 12, then the total number of bins is 84; if 24 bins per octave are used, then the total number of bins is 168; and so on. Fig.~\ref{fig:Exp1} shows the influence of changing the bin size on LogSpec. It can be seen that the rightmost one has a lower resolution. Throughout the experiment, a fixed hop size of 512 is used. For the MelSpec, a fixed STFT window size of 4,096 is used, and the number of Mel filter banks is varied from 128 to 2,048.

\subsection{Experiment 2: STFT window size}\label{sec:Exp2}
In the second experiment, we analyse the effect of the \emph{STFT window size}, together with varying bin size on the transcription accuracy for the MelSpec, LogSpec, and LinSpec based models. The STFT window size directly affects the frequency resolution of the spectrogram. A longer STFT windows size ensures a high \emph{frequency} resolution by sacrificing the \emph{time} resolution~\cite{wright1999short}. Because the STFT window size is kept constant in Experiment 1 (4,096), the time resolution in the previous experiment is fixed. In this experiment, however, the time resolution varies so that we can examine the effect of a high time resolution on transcription accuracy. For LinSpec and LogSpec, the number of bins is set to half of the window size of the short-time Fourier transform (STFT), so as to remove redundant information due to symmetry. For instance, when a STFT window size of 4,096 is used, 2,048 frequency bins are used to create the spectrograms, and so on. For MelSpec, a different number of Mel filter banks are explored for each varying value of STFT window size. Because the STFT window size for CQT cannot be adjusted independently (see Section~\ref{chap:result_exp2}), it is not studied in this experiment.  

\begin{figure}[htb]
	\centering
	\includegraphics[width=8.5cm]{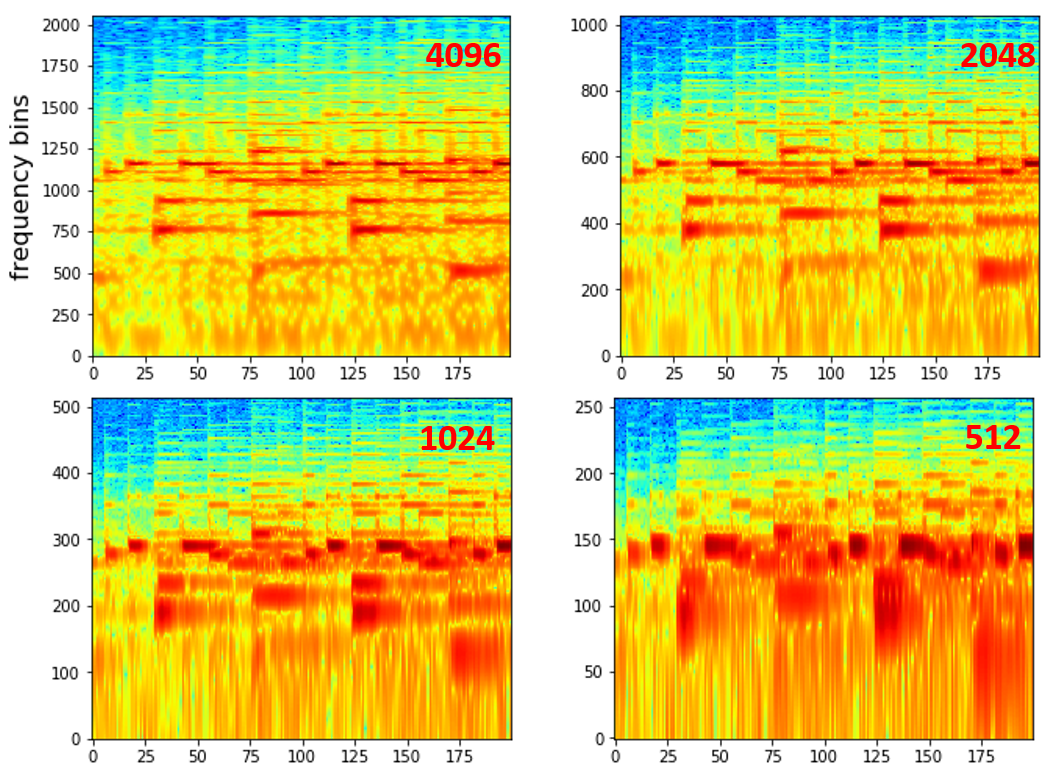}
	\caption{LogSpec of song ID:2303 with varying STFT window sizes of (4,096, 2,048, 1,024, and 512)}
	\label{fig:Exp2}
\end{figure}

\subsection{Experiment 3: Length of input frame size}\label{sec:Exp3}
In the previous two experiments, the input audio length is fixed at 16,384 samples. In this experiment, however, we study the effect of varying the input audio length from 0.14 to 0.45 seconds. All the other spectrogram parameters such as STFT window size (4,096), Mel filter banks (512), and number of bins per octave (72) are held constant.

\section{Results and Discussion}
\label{sec:results}

\begin{figure*}[htb]
	\centering
	\includegraphics[width=18cm]{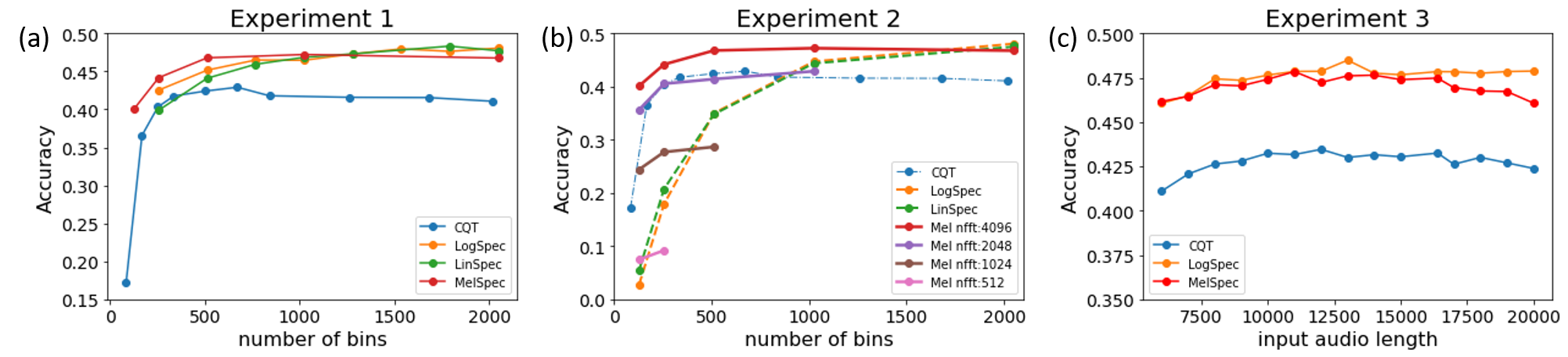}
	\caption{(a), (b), (c) shows the experimental results for Experiment 1, 2, and 3 respectively, in terms of transcription accuracy.}
	\label{fig:Linear_result}
\end{figure*}

Fig.~\ref{fig:Linear_result} shows the results for the three experiments, which are discussed in the following subsections.
\subsection{Experiment 1}
Intuitively, one might think that a higher spectrogram resolution would result in higher transcription accuracy.  Fig.~\ref{fig:Linear_result}(a), however, shows us that this is not necessarily true for all input representations, such as CQT. In the case of CQT, when the number of bins per octave is 12, the transcription accuracy is as low as 0.16. As the number of bins per octave is doubled to 24, the transcription accuracy increases dramatically to 0.36, and keeps improving slightly until the number of bins reaches 84. After that, the transcription accuracy for CQT deteriorates. When comparing CQT with other input representations, such as LogSpec or MelSpec, CQT-based models do not perform as well. This is consistent with the existing literature~\cite{Kelz2016OnTP, wang2018singing}. LinSpec and LogSpec have a comparable transcription accuracy, except when the number of bins is less than 768. In that case, LogSpec performs slightly better than the LinSpec. Since the bin spacings for LogSpec are similar to relationships between musical notes, LogSpec can still maintain useful information when the number of bins is low. When there are enough bins, the difference in transcription accuracy between LogSpec and LinSpec is less obvious because the large number of bins evens out the advantage that LogSpec has for low frequency signals. As for models based on MelSpec, these perform best when the number of Mel filter banks is 1,024. If more Mel filter banks are used, some of these will be empty. This is why MelSpec with too many Mel bins performs slightly worse than both log-frequency and linear-frequency spectrograms. This result also reveals that Mel filter banks can be considered an effective compression algorithm for spectrograms. While reducing the number of frequency bins they still maintain a relatively high transcription accuracy. Overall, the best performing input representation is LogSpec with an STFT window size of 4,096 and 2,048 bins.

\subsection{Experiment 2}
\label{chap:result_exp2}
Fig.~\ref{fig:Linear_result}(b) shows the influence of varying the STFT window size (or so-called n\_fft in some of the audio processing libraries) for the MelSpec representation. The transcription accuracy for a MelSpec with a specific window size is always lower than its log-frequency spectrogram counterpart with the same window size. The results further suggest that the Mel filter banks act as a spectrogram compression to reduce the number of frequency bins from the STFT result with minimum loss in information. The improvement in transcription accuracy becomes minimal when the STFT window size reaches 8,192; therefore, our experiment was stopped at 4,096. (Note that a window size of 4,096 results in only 2,048 non-redundant frequency bins.) For CQT, the only tunable parameter is the number of bins per octave when the minimum and maximum frequency are fixed, the number of bins (b) per octave affects the quality factor $Q=(2^{\frac{1}{b}}-1)^{-1}$, and hence the STFT window size $N_{k_{cq}}=   \text{ceil}{\left(\frac{s}{f_{k_{cq}}}\right)Q}$, where $s$ is the sampling rate and $f_{k_{cq}}$ the frequency for a specific bin.  Given that the number of bins per octave is the only tunable parameter for CQT, the CQT result here is same as Experiment 1; The overall best performance is still LogSpec with STFT window size 4,096 and 2,048 bins, closely matched by LinSpec. 

\subsection{Experiment 3}
The input audio length has only a minor effect on the music transcription accuracy. The length of the input audio affects the width (time-steps) of the spectrogram, and the precision of pitch labels. Fig.~\ref{fig:Linear} shows that a long input frame results in a reduction of accuracy. This could be because the input frame might contain multiple note transitions. Since the pitch label corresponds to the note appearing in the middle of the frame, longer frames might confuse the neural network. Fig.~\ref{fig:Linear_result}(c) shows that decreasing the audio clip length from 16,384 slightly improves transcription accuracy. The transcription accuracy, however, starts to decay once the audio clip is too short. The optimal performance is reached with an input length between 10,000 and 16,384 for LogSpec. 

\subsection{Comparison with state-of-the-art}
\label{chap:comapre}
Given the above results, the best input representations are compared with state-of-the-art models trained on the MusicNet database in Table~\ref{tab:result}. The comparison is made in terms of accuracy and error as defined in the \textbf{mir\_eval} library. The overall best result is found by using LogSpec and LinSpec based models with n\_fft 4,096, and 2,048 frequency bins. When comparing the results obtained from our simple but optimally-tuned networks to the state-of-the-art published results on the MusicNet dataset, only more complex CNN architectures~\cite{Trabelsi2017DeepCN} outperform the proposed system in terms of accuracy (other metrics are not given). In future work, we will explore the effects of input representations on more complex models and convolutional architectures. 
\begin{table}[htb]
	\centering
	\begin{tabular}{l|ccc} 
		\toprule
		Model & Precision & Accuracy & Error\\\hline
		MusicNet~\cite{Thickstun2017InvariancesAD}& 65.9 & $44.4$ & $61.8$\\
		Melodyne~\cite{Melodyne} & 58.8 & $41.0$ & $76.0$\\
		Deep Complex CNN~\cite{trabelsi2017deep} & 72.9 & N.A. & N.A. \\\hdashline
		LinSpec nfft: 4,096, bins: 512& 63.6 & $44.1$ & $60.7$\\
		LogSpec nfft: 4,096, bins: 512& 64.6 &$45.2$ & $59.3$ \\
		MelSpec nfft: 4,096, Mel bins: 512& 65.6&  $46.8$ & $58.5$\\\hdashline
		\pmb{LogSpec nfft: 4,096, bins: 2,048}& \pmb{66.6} & \pmb{$48.1$} & \pmb{$56.0$} \\
		CQT bins per ocatave: 72 & 61.0 & $42.9$ & $66.0$ \\
		MelSpec nfft: 4,096, Mel bins: 1,024& 65.8 & $47.2$ & $58.4$\\
		
		\bottomrule
	\end{tabular}
	\caption{Transcription results on the test set using the \texttt{mir\_eval} package~\cite{Raffel2014MIR_EVALAT}. The precision, accuracy, and error reported here are the frame precision, frame accuracy, and frame chroma total error as implemented in the \texttt{mir\_eval} package. Best model performance is in bold.} 
	\label{tab:result}
\end{table}

\section{Conclusions}
\label{sec:conclusions}
This study analyzes the importance of input representations on the accuracy of polyphonic, multi-instrument music transcription. We use our GPU based spectrogram toolkit, nnAudio~\cite{cheuk2019nnaudio}, to calculate on-the-fly spectrogram representations around 100 times faster than traditional libraries. In our experiments, we found that, by fixing the neural network architecture and only varying the input representation, we can improve the transcription accuracy by $8.33$\% (from $44.4$ to $48.1$), and reduce the error by $9.39$\% (from $61.8$ to $56.0$). When the resolution of the spectrogram is high enough (e.g., a greater number of frequency bins), we found that a log-frequency spectrogram representation ensures a better transcription result, at the expense of computational complexity. If the computational complexity is a concern, MelSpec can compress the spectrogram into a more compact representation while maintaining a relatively high transcription accuracy. A higher level spectral representation, CQT, shows relatively poor transcription accuracy. One reason might be the STFT window size being coupled with the number of frequency bins. Modifying the CQT algorithm such that the number of frequency bins can be varied without changing the STFT window size is one of our future research directions. Neural networks may be able to learn better from a lower level input representation, such as a log-frequency spectrogram or linear-frequency spectrogram. Future research will also explore the influence of additional input representations, such as a learned representation obtained by training the transcription end-to-end with raw audio as the input.

\section*{Acknowledgment}
We would like to thank the anonymous reviewers for their constructive reviews. This work is supported by a Singapore International Graduate Award (SINGA) provided by the Agency for Science, Technology and Research (A*STAR), under grant no. SING-2018-02-0204. Moreover, the equipment required to run the experiments is supported by SUTD-MIT IDC grant no. IDG31800103 and MOE Grant no. MOE2018-T2-2-161.

\bibliography{refs}
\bibliographystyle{IEEEtranN}

\vspace{12pt}

\end{document}